\documentclass[10pt]{article}
\usepackage{authblk}
\usepackage[utf8]{inputenc}
\usepackage[T1]{fontenc}
\usepackage{graphicx}
\usepackage{amsmath}
\usepackage{amssymb}
\usepackage{hyperref}
\usepackage[margin=30mm]{geometry}
\usepackage{parskip}
\usepackage{bm}

\newcounter{cntAlg}
\newcounter{cntStrAlg}[cntAlg]
\newlength{\algCaptUpVSkip}
\newlength{\algCaptDnVSkip}
\newlength{\algCaptHSkip}
\newlength{\algCaptAlg}
\newlength{\algStringNumWidth}
\newlength{\algStringAfterNumOffs}
\newlength{\algTabulator}
\newlength{\textAwidth}
\DeclareDocumentCommand{\algLine}{O{0} O{} m}{%
    \setlength{\textAwidth}{\textwidth}
    \noindent\parbox{\algStringNumWidth}{\refstepcounter{cntStrAlg}%
        \hfill\small\thecntStrAlg:\hspace*{\algStringAfterNumOffs}%
        \IfNoValueTF{#2}%
            {}%
            {\label{#2}}%
	}
        \IfNoValueTF{#1}%
            {}%
            {\hspace*{\dimexpr(#1\algTabulator)\relax}}%
	\parbox[t]{\dimexpr(\textAwidth-#1\algTabulator-\algStringNumWidth)\relax}{#3}%
}
\DeclareDocumentCommand{\algLineDouble}{O{0} O{} m m}{%
    \setlength{\textAwidth}{82.5mm}
    \noindent\parbox{\algStringNumWidth}{\refstepcounter{cntStrAlg}%
       \hfill\small\thecntStrAlg:\hspace*{\algStringAfterNumOffs}%
        \IfNoValueTF{#2}%
            {}%
            {\label{#2}}%
	}
        \IfNoValueTF{#1}%
            {}%
            {\hspace*{\dimexpr(#1\algTabulator)\relax}}%
	\parbox[t]{\dimexpr(\textAwidth-#1\algTabulator-\algStringNumWidth)\relax}{#3\phantom{!}}%
	\hfill
	\parbox{\algStringNumWidth}{\hfill\small\thecntStrAlg:\hspace*{\algStringAfterNumOffs}%
        \IfNoValueTF{#2}%
        {}%
        {\label{#2}}%
    }
        \IfNoValueTF{#1}%
            {}%
            {\hspace*{\dimexpr(#1\algTabulator)\relax}}%
	\parbox[t]{\dimexpr(\textAwidth-#1\algTabulator-\algStringNumWidth)\relax}{#4\phantom{!}}%
    \setlength{\textAwidth}{\textwidth}
}
\title{Two decades of algorithmic Feynman integral reduction}
\author[1,3]{Alexander Smirnov}
\author[2,3]{Vladimir Smirnov}
\affil[1]{Research Computing Center, Moscow State University, Moscow, Russia}
\affil[2]{Skobeltsyn Institute of Nuclear Physics of Moscow State University, Moscow, Russia}
\affil[3]{Moscow Center for Fundamental and Applied Mathematics, Moscow State University, Russia}
\date{\today}
\begin{document}
\maketitle

\begin{abstract}
We present a historiographical review of algorithms and computer codes developed for solving integration-by-parts relations for Feynman integrals. This procedure is one of the key steps in the evaluation of Feynman integrals, since it enables to express
integrals belonging to a given family as linear combinations of master integrals. In this review, we restrict ourselves to considering general algorithms which can, in principle, be applied to any family of Feynman integrals.
\end{abstract}

\tableofcontents
 
\section{Introduction}

When performing evaluations in perturbative quantum field theory, one needs to compute 
dimensionally regularized Feynman integrals from a family of scalar integrals associated with a given diagram.
The integrals have the following form
\begin{equation}
    I_{a_1,\ldots,a_n} =  
  \int\ldots\int \frac{1}{P_1^{a_1}\dots P_{n}^{a_{n}}} 
   \prod_{i=1}^h {\rm d}^d k_i,
\label{FI}  
\end{equation}
where $d=4-2\varepsilon$ is the parameter of dimensional regularization, $P_i$ are 
denominators of propagators
which are linear combinations of Lorentz scalar products $k_i\cdot k_j$ and $k_i\cdot p_l$
where $p_l$ are external momenta and $k_i$ are $h$ loop momenta.
The $n$ integer numbers $a_i$ are called {\em indices}. 
After applying Feynman rules to a family of integrals, one obtains integrals with Lorentz
indices. Then the usual procedure is to express them in terms of scalar integrals (\ref{FI}),
where the indices can take, generally speaking, any integer values.
Feynman integrals are rational functions of the regularization parameter $d$ with poles 
arising from ultraviolet, infrared and collinear divergences.

The number of resulting integrals can be very large but
in the early days of perturbative quantum field theory it was necessary to evaluate
every appearing Feynman integral by some method.
The situation has significantly changed after the groundbreaking discovery
of Chetyrkin and Tkachov \cite{IBP1}, who suggested to apply  
so-called integration by parts relations (IBP) in order to express any given
Feynman integral as a linear combination of a finite number of Feynman integrals.
These relations are derived from equations
\begin{equation}
\int\ldots\int 
\left(\frac{\partial}{\partial k_j}\cdot r\right) \left(  
  \frac{1}{P_1^{a_1}\dots P_{n}^{a_{n}}}
\right)  
   \prod_{i=1}^h {\rm d}^d k_i =0,
\label{IBP}  
\end{equation}
where $r$ is a loop momentum $k_j$ or an external momentum $p_l$ and $j=1,\ldots,h$.
After the differentiation, resulting scalar products $k_j\cdot k_{j'}$ and $k_j\cdot p_l$  
are expressed in terms of the factors in the denominator, and one obtains 
a system of equations which are called IBP relations. The dependence on $d$ appears due to 
the prescription $\frac{\partial}{\partial k_j}\cdot k_j=d$ of dimensional regularization.
When solving IBP relations Feynman integrals are considered as algebraic objects. However, when evaluating
a given Feynman integrals analytically, one usually applies Feynman parametric representation
given by a multifold parametric integral, where $d$ enters as a complex parameter in exponents
of some polynomials.

The very first impressive application of the IBP relations was immediately presented in \cite{IBP1},
where an algorithm to evaluate massless three-loop propagator integrals was described.
Later this algorithm was implemented as the program {\tt MINCER} \cite{MINCER1,MINCER2} 
(implemented in {\tt FORM} \cite{FORM})  
to evaluate this class of Feynman integrals. It was used in numerous applications.
We will mention below only one similar result 
because we are oriented at general algorithms and corresponding computer codes
which can, in principle, be applied to any family of Feynman integrals.

The IBP relations following from (\ref{IBP}) can be written down as operator 
equations with the help of \textit{annihilation/creation} operators
\begin{equation} 
{\bf i}^{\pm}  I_{a_1,\ldots,a_n} = I_{a_1,\ldots,a_{i-1},a_{i}\pm 1,a_{i+1},\ldots a_n}\,. \nonumber
\end{equation}
Starting from these operator equations one obtains an infinite set of equations for functions
of integer variables $a_i$. 
From the mathematical point of view, the IBP reduction is solving a huge sparse system of linear equations with polynomial coefficients.

To be more exact, when one fixes a subset of unknown Feynman integrals (that is, the possible values of indices), and writes down all possible IBP relations connecting them one obtains a sparse system of linear equations with polynomial coefficients. This system can be solved reducing all unknowns to integrals that are smallest under some chosen ordering.
It was observed long ago that if one keeps increasing the number of possible indices, and thus the number of integrals involved, the number to which the others are reduced eventually stabilizes. This number remains constant no matter how much the indices are increased.
Such irreducible integrals are called {\em master} integrals. 
This, however, is not a strict mathematical definition, and a more rigorous formulation is beyond the scope of this paper.
It was proved in \cite{Smirnov:2010hn} that the number of master integrals is always finite. Nevertheless, this fact does not help in calculations since the proof is purely theoretical.

In practice, one typically considers a sufficiently large subset of Feynman integrals from a given family., for example, all integrals for which the inequality $\sum\limits_{i=1}^n |a_i|\leq M$ holds.  One can refer to these integrals as {\em seeded} Intergals.
One then writes down all the IBP relations that involve only the integrals in this subset and solves the resulting system.
One introduces an ordering when solving such a system: integrals with small values of $|a_i|$ are naturally considered to be on the right-hand side of the corresponding
solutions. 
Following Laporta's presentation of the first detailed version of such an algorithm \cite{Laporta}, these methods became widely known as "Laporta algorithms."

However, the strategy formulated above is far from being optimal.
It turns out that in advanced strategies, the notion of {\em sector} is used.
A sector of integrals of a given family (\ref{IBP}) is determined by a set of positive indices $a_i$
in (\ref{FI}), while the rest of indices are non-positive.
It is enough to construct a reduction procedure in each sector. Then
a Feynman integral from a given sector can be reduced to master integrals in the same
sector up to a linear combination of integrals of {\em lower} sectors, i.e. for which
some of the positive indices become non-positive. There is a variety of ways
to construct the reduction in a given sector presented in available algorithms.

The goal of this paper is to present a historiographical review of algorithms and computer programs
to solve IBP relations with explanations of their essential features.
We follow the chronological order also because this is a story about competition
in the field of solving IBP relations. It turns out that this competition was 
fruitful since existing algorithms were taken into account when constructing new
algorithms and/or new versions of existing algorithms.
We do not refer at all to papers with applications of these algorithms and computer codes
in practical calculations because the number of corresponding applications is exceedingly large. However, they can easily be found by studying citations to a given algorithm.
By examining the applications and focusing on the most important ones, a researcher can decide which algorithm is more suitable for a given problem. 
 
\section{AIR}
 
The first computer implementation of the Laporta algorithm was created by Anastasiou 
and Lazopoulos \cite{Anastasiou:2004vj}
and called Automatic Integral Reduction ({\tt AIR}).
To define enlarging sets of seeded integrals, the following functions of integrals
of a given family $I(\{a_i\})$ are introduced:
\[
N_{\rm prop} = \sum_i \Theta(a_i), 
\qquad
N_+ = \sum_i \Theta(a_i)\,(a_i - 1), 
\qquad
N_- = -\sum_i \Theta(-a_i)\,a_i ,
\]
where $\Theta(x)=1$ for $x>0$ and $0$ otherwise.  
At each elimination step, the most ``complicated'' integral (highest $N_{\rm prop}$, $N_+$, $N_-$ in lexicographic order) is isolated and expressed in terms of simpler ones.
The user specifies ranges of $(N_{\rm prop}, N_+, N_-)$ and template IBP/LI relations for a given family.
Here and below by LI we mean so-called Lorentz-invariance identities \cite{LI}, which are used together
with IBP relations.
This allows AIR to generate
explicit ``seed equations'' for concrete sets of indices $\{a_i\}$.  
The resulting equations are sorted and used in the elimination procedure.

The main reduction cycle of {\tt AIR} consists of the following steps:
\begin{itemize}
  \item selecting a seed integral;  
  \item generating IBP/LI equations;  
  \item ordering terms by complexity and isolating the hardest integral;  
  \item solving 
         in terms of simpler ones;  
  \item performing back-substitution into previously derived equations.  
\end{itemize}
This iterative process gradually reduces all integrals to master integrals.

To control the growth of intermediate algebraic expressions, AIR employs also two masking strategies.
Equations are stored in a file database, where each one is
associated with its target (isolated) integral.  
Auxiliary index files keep track of where each integral appears.  
At the end, masked objects are re-expanded through nested substitutions and 
any given integral $I$ is expressed as
$I = \sum_j c_j\left(\{s_l\}, d\right)M_j$, where $M_j$ are master integrals
and coefficients $c_j$ are rational functions of
kinematic invariants $s_l$, particle masses and space-time dimension.

\section{FIRE 1--3}

Before the public release of {\tt FIRE} (Feynman Integrals REduction), our work was focused on building prototype systems referred as {\tt FIRE1} and {\tt FIRE2} (written in {\tt Wolfram Mathematica}), which were never made public.

The initial formulation of {\tt FIRE} was strongly influenced by our attempt to apply Gr\"obner basis techniques to the reduction of Feynman integrals \cite{Smirnov:2005ky,Smirnov:2006tz}.
The IBP relations can be represented as polynomial relations in shift operators acting on indices of Feynman integrals. By adapting and modifying the Buchberger algorithm~\cite{Buch}, the Gr\"obner-type bases in this operator algebra were constructed. This perspective provided a systematic framework for solving the reduction problem: 
an arbitrary integral from a given family can be reduced to
a linear combination of a smaller set of master integrals by repeated application of these basis relations.
In {\tt FIRE1} the user had to provide manually \textit{boundary conditions}, which characterize index regions where integrals take zero values, thus allowing the algorithm to terminate effectively.

The version {\tt FIRE2} was an update of {\tt FIRE1}
that incorporated several new features
such as symmetry relations and internal optimizations. 
Still at this point the authors came to a conclusion that it becomes really
impossible to construct Gr\"obner bases in some sectors. 
Therefore, to move forward, the Laporta algorithm had to be implemented in {\tt FIRE}.
The first public version of {\tt FIRE} named {\tt FIRE3} still written in {\tt Wolfram Mathematica}
and having a hybrid approach, combining Gr\"obner-basis with Laporta elimination, was published in 2008
\cite{Smirnov:2008iw}.
The version {\tt FIRE3} possessed the following features:

\begin{itemize}
 \item Laporta mode: a classical IBP-based elimination in sectors, used when no s-basis is available for a given sector;
 \item  s-basis mode: in sectors where an s-basis has been constructed, any non-master integral can be reduced to lower ones in (essentially) constant time by symbolic substitution, avoiding a full system solve;
 \item region-bases: a feature allowing to treat shifted indices.
\end{itemize}

To bring in a proper formalism and to define master-integrals in a mathematical manner, the notion of a global integral ordering was refined in {\tt FIRE3}.
A global ordering is defined first by diagram number, then by sector, and then within the sector by a linear ordering on the shift vector representation. Sectors are defined by sign patterns of indices (positive vs non-positive). Regions generalize sectors allowing some indices to be ``passive'' (i.e. not fixed sign). The presence of region definitions requires that the ordering be adjusted so that integrals lying in the lower regions appear ``lower'' in the partial order, and that reduction steps do not violate region consistency. With these definitions {\tt FIRE3} formalizes a concept: a \textit {proper expression} for an integral $I$ is an identity expressing $I$ as a linear combination of integrals strictly lower in the ordering. The core algorithm ensures that eventually all integrals of interest possess a proper expression and at this stage
the backward substitution reduces them
to masters integrals. The master integrals can now be defined like integrals for which no a proper expression can exist.

To manage the growth of the number
of expressions, {\tt FIRE3} uses {\em masking} techniques (analogous to those in {\tt AIR}).
When an IBP relation involves integrals from lower sectors, the parts corresponding to lower-sector integrals are masked (not substituted immediately). This
is called {\em tail-masking}. The substitution of such expressions is delayed. The tail-masking concept turned out to be important for future reduction development, especially within the {\tt FIRE} framework working unlike other reduction programs in two steps: first, top-bottom reduction with tails being masked and then, a bottom-top backward substitution.

In analogy to approaches proposed by Lee \cite{Lee:2012cn}, {\tt FIRE3} could also eliminate
certain IBP relations deemed redundant and only generate a subset sufficient to express most integrals in a sector. This reduces the number of IBPs one must consider in system solving.

While still being written in {\tt Wolfram Mathematica,} {\tt FIRE3} introduced such tools as {\tt QLink}, an interface to the QDBM disk-based database for managing large algebraic storage  
within Mathematica, and {\tt FLink}, a bridge to the Fermat program, which can perform fast polynomial arithmetic and substitution outside Mathematica.

\section{Reduze}

The next implementation of the Laporta algorithm was presented by Studerus \cite{Studerus:2009ye} in $2009$, where some notation was introduced to be used in future papers too.

Let $T_t$ be the set of integrals of a given family associated with integrals with $t$ indices.
A sub-sector $T_{t-1}$ of a sector $T_{t}$ is a sector where one propagator is removed.
There are in general $t$ different sub-sectors for a
sector $T_{t}$. The sub-sector tree of a sector $T_t$ is the set of all sub-sectors
of $T_t$ and, recursively, all sub-sectors of these sub-sectors.
In other articles \textit{sub-sectors} can also be named as \textit{lower sectors}.

The following additional notation is also introduced in \cite{Studerus:2009ye}.
To every $t$-propagator sector $T_t$ with propagators $P_{j_1}$, $\ldots$, $P_{j_t}$
belongs a infinite set of $d$-dimensionally regularized $l$-loop integrals, which all share the same 
propagators in the denominator of the integrand. They have the generic form
\begin{equation}\label{eq:dim_reg_int}
\int \mbox{d}^d k_1 \ldots \int \mbox{d}^d k_l ~
     \frac{P_{j_{t+1}}^{s_1} \ldots P_{j_n}^{s_{n-t}}}{P_{j_1}^{r_1} \ldots P_{j_t}^{r_t}}
\end{equation}
with integer exponents $r_i \geq 1$ and $s_i \geq 0$.
Then, one defines $r=\sum_{i=1}^t r_i \geq t$ as the sum of the indices propagators in the denominator and
$s=\sum_{i=1}^{n-t} s_i \geq 0$ as the sum of the indices of propagators in the numerator.
These two numbers $r$ and $s$ identify the complexity of integrals of a given sector. But since $r$ is obviously not smaller than the number of positive indices, nowadays a shifted formula can be used with $r=\sum_{i=1}^t (r_i - 1) \geq t$, so that $r$ and $t$ together can be called a complexity of the shift from the corner of the sector.

To reduce the integrals of a sector up to $r=r_{\rm{max}}$ and $s=s_{\rm{max}}$, it is necessary to solve
the homogeneous system of linear equations, which follow from the corresponding IBP and/or LI identities. Then, one obtains a solution in terms of some integrals of a given sector and integrals of lower sectors. 
Afterwards, the procedure continues in these lower sectors and
finally a solution in terms of true master integrals is obtained.

The reduction scheme of {\tt Reduze} is as follows.
For the reduction of a sector and its sub-sectors, {\tt Reduze} first constructs the full sub-sector tree. It begins by reducing the lowest sub-sectors (with the fewest propagators), substitutes their results into higher sectors and continues iteratively until the target sector with the largest number of propagators is reduced. For a single-sector reduction, {\tt Reduze} generates the corresponding equations and inserts available sub-sector results from the default directory.

Since {\tt Reduze} cannot reduce a single integral directly, it uses all integrals within a user-defined range of $r$ and $s$ to form and reduce the full equation system.
Because these systems can be very large, {\tt Reduze} automatically divides them into smaller subsets. Equations are sorted by the complexity of their most complicated integral, split into user-defined chunks, and stored in temporary files.
The reduction proceeds sequentially: the simplest subset is reduced first, its results are inserted into all others, and the process repeats for each subsequent subset. In the end, all results are combined into a single output file.

The approach of {\tt Reduze} here differs from the approach of {\tt FIRE} since it uses one bottom-top pass when solving equations.

\section{Reduze 2}

The second version of {\tt Reduze} \cite{vonManteuffel:2012np},
presented by von Manteuffel and Studerus in $2012$, was 
a considerable revision
and extension of {\tt Reduze 1}, rearchitected to support the distributed algorithmic framework. The main feature was built around the
distributed (MPI‐based) approach to reduction, supporting
distributed reduction of a single family of integrals (or, a single sector) across multiple processor cores using MPI hence using multiple computers for the same task. It also added a support to a
parallel reduction of different families and a load balancer to properly distribute jobs between nodes.

This is a summary of other new features:
\begin{itemize}
\item graph and matroid–based algorithms for sector/integral relations. 
A key improvement is the use of fast graph- and matroid-based methods to identify equivalences (relations) among sectors (families) and integrals, including those across different families, and to match diagrams to canonical sectors.
These algorithms help to reduce redundancy and simplify the reduction workload;

\item improved handling of sector symmetries and relations. 
{\tt Reduze 2} enhances the detection and utilization of sector relations (between sectors of the same or different families) and sector symmetries (automorphisms), beyond the symmetry handling in the first version;

\item matching of diagrams to canonical sectors. 
{\tt Reduze 2} implements routines to match a given Feynman diagram (with its labeling) to the canonical representative sectors (via mapping or shifts);

\item differential equations for master integrals support. 
{\tt Reduze 2} includes features to derive differential equations for master integrals (in the kinematic variables) as a built-in option;

\item interference-term reduction. 
The program can reduce interference terms (e.g. integrals appearing in interference of amplitudes) directly.

\end{itemize}
 
\section{LiteRed}

In 2012 Lee introduced {\tt LiteRed}, a Mathematica package aimed at automating the heuristic derivation of symbolic IBP reduction rules (rather than repeatedly solving linear systems) \cite{Lee:2013mka,Lee:2012cn}.
{\tt LiteRed} departs, to some extent, from the traditional Laporta-style approach of solving large IBP systems anew for each integral. Instead, its goal is to precompute reduction rules (in a sector-by-sector manner) 
that can later be applied directly to any integral in that family.
Once the symbolic rules are found, reduction becomes a matter of rule application, which is extremely fast and lightweight. As the author states, symbolic rules are compact and can be stored for reuse, 
avoiding repeated costly system solving. Thus, in a way {\tt LiteRed} follows the ideas of {\tt FIRE1-2}, but the author developed his own way of precomputing reduction rules not related to Gr\"obner bases.

However, the trade-off is that the search for symbolic rules is heuristic and not guaranteed to succeed in all cases. The algorithm may fail or stall in particularly challenging sectors, and success often depends on ordering choices or the particular set of irreducible numerators chosen.

{\tt LiteRed}’s workflow broadly splits into two phases: rule search (preprocessing) and rule application (actual reduction). Precomputation consists of the following steps:
\begin{itemize}
 \item the user defines a basis for integrals (i.e.\ denominators and optional numerators) via a ``NewBasis'' command. {\tt LiteRed} checks linear independence and completeness of the basis representation in terms of scalar products;
 \item IBP and Lorentz-invariance identities are generated via {\tt GenerateIBP}, producing symbolic relations in shift-operator form;
 \item sectors are analyzed via {\tt AnalyzeSectors}: zero (scaleless) sectors are identified, and ``simple'' sectors are recognized;
 \item symmetries between sectors are detected through {\tt FindSymmetries}, which produces mappings between equivalent or isomorphic sectors, generating substitution rules to reduce duplication;
 \item the central routine {\tt SolvejSector} attempts to construct a full set of symbolic reduction rules in a given sector (for each unique sector after symmetry factoring). The outcome is a set of reduction rules ``jRules'', which express non-master integrals in terms of simpler ones.
\end{itemize}

The sector solving approach of {\tt LiteRed} turned out to be more effective than the one in {\tt FIRE1-2}.
Let us note that some important information revealed with {\tt LiteRed} is used in the next versions of {\tt FIRE} (see below).

\section{FIRE4}

Published in $2013$ it was the first version \cite{Smirnov:2013dia} of {\tt FIRE} that 
featured reduction in {\tt C++} providing a serious
performance upgrade with a direct access to databases and {\tt Fermat} algebraic evaluations.
Apart of that it has the following features:
\begin{itemize}
  \item integration with ``tsort'' using an
 algorithm (based on Feynman parameters and described by Pak 
  \cite{Pak:2011xt}) to search for equivalences between integrals and detect additional linear relations among integrals (beyond IBP) to reduce the master set;
  \item use of symmetries and Lorentz-invariance identities to derive further relations among master integrals, which helps to eliminate spurious masters;
  \item integration with {\tt LiteRed} --- the new version of {\tt FIRE} was able to provide input for {\tt LiteRed} and to use its output for the {\tt C++} reduction both handling symmetry relations and reduction rules;
  \item the command ``MakeMaster'' to allow the user to assign priorities to particular integrals to be chosen as masters in ambiguous sectors.
\end{itemize}

In the case of {\tt LiteRed} integration it turned out to be always beneficial to use {\tt LiteRed} symmetries between sectors. {\tt FIRE4} could also use {\tt LiteRed} reductions rules resorting to internal reduction methods in the missing sectors in case {\tt LiteRed} does not provide full coverage. However, the further usage has shown that it is beneficial either to build {\tt LiteRed} reduction rules everywhere or not to use them at all, since the rules might greatly increase the complexity of integrals to be reduced in lower sectors and when
the rules are missing there, the reduction becomes more complex than the original one without {\tt LiteRed} rules. Let us mention that {\tt LiteRed} was used to construct {\tt FORCER} \cite{Ruijl:2017cxj}, i.e.
a four-loop variant of the {\tt MINCER} code \cite{MINCER1,MINCER2}.
Here, {\tt LiteRed} provided a set of symbolic reduction rules, though these may not be optimal.
These rules were then essentially optimized for most complicated
four-loop diagrams.
  
\section{FIRE5}
 
Then {\tt FIRE5} \cite{Smirnov:2014hma} was released in $2014$. This was a major milestone in the FIRE lineage: the core reduction engine was rewritten in {\tt C++}, while Mathematica remained as the user-facing frontend and orchestration layer. This structural shift allowed to overcome many performance and scalability limitations of a pure Mathematica implementation. The whole scheme introduced at this stage
was the following (and remained true for more recent versions):
\begin{itemize}
 \item {\tt Mathematica} frontend handles problem setup (diagram definition, generation of IBP relations, symmetry declarations, sector definitions, boundary conditions), programming interface, and final result collection;
 \item {\tt C++} backend performs the heavy reduction steps: equation solving, substitution into integrals, table building;
 \item interfacing between them
is handled via files/tables produced by {\tt Mathematica} that the {\tt C++} engine consumes, and result tables emitted by {\tt C++} that can be loaded back into {\tt Mathematica};
 \item the design is backward-compatible: reductions produced by {\tt FIRE5} can be imported into earlier {\tt FIRE} versions, and the same user workflow (e.g. {\tt Mathematica} commands like {\tt Burn[]}, {\tt LoadTables}) remains available.
\end{itemize}

By moving the bottleneck parts to {\tt C++}, the version {\tt FIRE5} achieved orders-of-magnitude improvements in performance for nontrivial reductions.

The {\tt C++} reduction implementation brought the following features:
\begin{itemize}
\item multi-threading/parallelism: {\tt FIRE5} supports parallel reduction tasks over multiple cores;
\item {\tt Fermat} integration: for symbolic polynomial arithmetic (simplification, GCDs, rational function operations) the {\tt C++} engine delegates to the external {\tt Fermat} program;
\item KyotoCabinet database backend: used for efficient on-disk storage of tables and sparse relations, allowing large intermediate data beyond in-memory capacity; the database settings can adjust how aggressively it caches in memory or spills to disk depending on available RAM, to optimize performance;
\item compression with Snappy: intermediate tables and data structures are compressed to reduce disk space and I/O load.
\end{itemize}

\section{Kira}
 
The first version of {\tt Kira} \cite{Maierhofer:2017gsa} by Maierh\"ofer, Usovitsch and Uwer published in $2017$ was designed to handle problems with multiple mass and kinematic scales, mitigating symbolic blowup in intermediate expressions.
{\tt Kira} is written in {\tt C++}, targeting 64-bit Linux systems, with support for multi-core parallelism.
It uses external algebraic support from {\tt Fermat} (via a ``gateToFermat'' interface) and {\tt GiNaC}, and employs libraries like yaml-cpp, zlib, and SQLite3 for configuration and storage. 

The reduction workflow is as follows:
\begin{itemize}
  \item generate a system of linear relations (IBP $+$ Lorentz invariance $+$ symmetry relations) among integrals in chosen seed sets;   
  \item perform a preprocessing elimination step using  modular arithmetic  to detect and remove linearly dependent equations before symbolic elimination. This step also helps to identify master integrals early;   
  \item using the remaining independent equations, perform forward elimination (triangularization) according to a chosen ordering of integrals and equations;    
  \item execute a back-substitution phase to express dependent integrals in terms of master integrals; the algorithm is optimized to delay full expansions and reduce intermediate expression swell;   
  \item post-process result tables, output mappings to master integrals, and provide user interfaces for queries.  
\end{itemize}

A given set of Feynman integrals are ordered lexicographically by a tuple involving (topology ID, sector ID, 
the sum of positive indices $r$, the sum of negative indices $s$, and detailed index vectors). Equations are sorted by their leading integral (largest in the ordering), then by equation length.  
Symmetry relations among integrals and zero (trivial) sectors are detected and exploited to reduce system size.   
 
The modular-arithmetic preprocessing is the principal novelty: it avoids needless symbolic operations on redundant equations and thus alleviates memory and time overheads especially in multi-scale settings.
Benchmark comparisons to {\tt Reduze 2} \cite{vonManteuffel:2012np} and {\tt FIRE5} \cite{Smirnov:2014hma} showed that {\tt Kira} is highly competitive, sometimes outperforming them in challenging multi-scale reductions.
Nevertheless, for extremely large systems the memory and CPU demands remain significant; the success of {\tt Kira} depends on effective seeding, symmetry detection, and the ability to eliminate redundant relations early.  
 
\section{FIRE6}

The release of {\tt FIRE6} \cite{Smirnov:2019qkx} in $2019$ marked a major revision of the {\tt FIRE} framework, integrating modular arithmetic into its {\tt C++} core and introducing large-scale parallel capabilities via MPI. While {\tt FIRE5} already provided a fast {\tt C++} engine, {\tt FIRE6} significantly extended both its computational reach and stability for extremely large reductions.

The most important innovation was \textit{the modular arithmetic reduction mode}, inspired in part by {\tt Kira}’s finite-field preprocessing. Instead of performing symbolic elimination over rationals, {\tt FIRE6} presents a mode to run multiple reductions for many integer substitutions of variables modulo large primes and later reconstructs the rational coefficients. This method, already explored in {\tt Finred} (see a short description in Conclusion) and {\tt Kira}, was implemented here using MPI parallelization across compute nodes to run reductions in parallel over distinct primes.

Key features of {\tt FIRE6} include:
\begin{itemize}
\item full support for distributed reduction across multiple computers and nodes (MPI mode), allowing scalability up to thousands of cores;
\item implementation of modular arithmetic reconstruction to mitigate coefficient blowup and accelerate large-scale reductions;
\item automatic recovery from system crashes --- reduction tasks can resume from checkpoints instead of restarting;
\item integration of internal sector symmetries and rule generation from {\tt LiteRed}, improving elimination efficiency and master-integral identification;
\item upgraded I/O and compression subsystems (using {\tt Snappy}, {\tt ZStandard}, {\tt KyotoCabinet}), improving disk efficiency and database handling;
\item enhanced {\tt C++11} compliance, improved documentation, and {\tt doxygen}-based internal API descriptions for contributors.
\end{itemize}

Conceptually, {\tt FIRE6} incorporates {\tt Kira}’s finite-field strategy but remains distinct in its execution model: {\tt Kira} uses modular arithmetic primarily for dependency detection and reconstruction, while {\tt FIRE6} applies modular arithmetic directly as the main reduction engine.
Its implementation is therefore more deeply parallelized and designed for heterogeneous environments --- from desktops to supercomputers.

However, {\tt FIRE6} had a significant disadvantage --- the reconstruction could be performed only within {\tt Wolfram Mathematica}. (This was later improved in {\tt FIRE7}). Nevertheless,
there was a number of successful projects, where the reduction was carried out using
supercomputers with subsequent reconstruction of coefficients.
 
\section{Kira2} 

The main novelity that was introduced in the second version of {\tt Kira} \cite{Klappert:2020nbg} presented by Klappert, Lange, Maierh\"ofer and Usovitsch in $2020$ was Finite field reduction using integrated FireFly package (first published as a standalone code \cite{Klappert:2019emp,Klappert:2021}) with a built in MPI support for distributing reduction jobs. Unlike {\tt FIRE6}, {\tt Kira} came to a complete solution making the reconstruction possible without proprietary software such as {\tt Wolfram Mathematica}. The {\tt FireFly} usage also enables {\tt Kira} to separate
known denominators (prefactors) from coefficients prior to interpolation, thereby simplifying the interpolation task.

{\tt Kira 2.0} significantly extended the flexibility, scalability, and efficiency of Feynman integral reductions.
Apart of that {\tt Kira2} provides
a number of new features:
\begin{itemize}
\item user-defined systems of equations --- an improved interface for supplying one’s own linear systems
(e.g. not necessarily standard IBP systems). Support for multiple input files, compressed files, 
on-the-fly solving, and custom integral weights. Also it allows more flexible definitions of propagators (e.g. combinations with Feynman parameters, bilinear forms);

\item iterative reduction --- automating splitting reductions over master integrals or sectors (setting other masters or sectors to zero), reducing memory footprint by iterating over subsets. Works both with algebraic and finite-field modes.
Also {\tt Kira2} allows marking entire sectors as trivial (all integrals zero) so that they are excluded from reduction;

\item master equations --- allows treating linear combinations of integrals as master integrals
by adding equations equating them to such combinations.
Useful for choosing convenient bases, e.g. for differential equations;

\item preferred-master sector inclusion. If a master integral lies in a sector not explicitly requested, Kira ensures that its sector is still generated to support that master integral. This helps find ``magic'' relations via higher; sectors.

\item export and re-import of reduction rules. With \texttt{kira2file} one can export reduction rules in a format compatible with user-defined input systems; results may also be exported to Mathematica or FORM via \texttt{kira2math}, \texttt{kira2form}.

\end{itemize}

\section{FIRE6.5}

In $2023$ the intermediate release {\tt FIRE6.5} \cite{Smirnov:2023yhb} introduced a new library, {\tt FUEL} \cite{FUEL}, which makes it possible to simplify
algebraic coefficients not with a single Computer Algebra System (CAS) but allows integration with multiple simplifier backends. {\tt FUEL}
provides a unified interface to multiple simplifier backends (e.g. {\tt FLINT}, 
{\tt Symbolica} \cite{SYMBOLICA} and {\tt Fermat}). The open-source {\tt FLINT} \cite{FLINT} backend was newly added to this release and
often outperforms Fermat in problems involving many kinematic variables. The {\tt Symbolica} simplifier (by Ruijl) is also integrated as a potential successor to {\tt FORM}.
Thus, {\tt FIRE6.5} becomes more flexible: the user may choose which simplifier to use, and the core engine is no longer bound to one CAS. To maintain backward compatibility, the default behavior remains Fermat.

\section{Kira3} 

{\tt Kira3} \cite{Lange:2025fba} published in $2025$ by Lange, Usovitsch and Wu, introduced optimized seeding and equation selection algorithms, significantly improving performance for multi-loop and multi-scale problems.
The following features can be marked as most important:

\begin{itemize}
\item
optimized seeding strategy --- the selection of seeds is improved. {\tt Kira3} inherits bounds $(r_{\rm{max}}, s_{\rm{max}})$ (see eq.~(\ref{eq:dim_reg_int})) more flexibly across subsectors, and allows {\em truncation} of seeds so that only a subset of seeds are generated that suffice for full reduction;
\item
better equation selection --- after generating linearly independent IBP equations, {\tt Kira3} employs a refined selection algorithm: it performs a numerical finite-field check post forward elimination to detect hidden zeros and drop irrelevant equations. This reduces the number of selected equations and, thus, speeds up the solving step;
\item support for extra relations --- users can now include additional linear relations (e.g. from nontrivial symmetries or kinematic constraints) in the standard reduction workflow by specifying {\tt extra\_relations} in the job file;
\item numerical sampling mode --- the user may provide finite-field sample points (via {\tt numerical\_points}) to drive reductions numerically, possibly bypassing full analytic expressions or guiding the interpolation step. {\tt Kira} ensures that the seed generation is consistent with the sampled phase space to avoid loss of information;
\item symbolic IBP reductions --- {\tt Kira3} can treat one or more propagator powers symbolically (i.e. as formal symbols) and derive recurrence (lowering/raising)  relations by Laporta’s approach. This allows to generate analytic recursion relations in favorable cases;
\item master basis checking --- with the option {\tt check\_masters}, {\tt Kira3} verifies whether a user-specified preferred master integral basis is sufficient. If not, it reports missing or superfluous masters and aborts.
\end{itemize}
  
\section{FIRE7}

The release of {\tt FIRE7} \cite{Smirnov:2025prc} marks a significant evolution of the {\tt FIRE} program, focused on optimizing the performance of IBP reduction under modular arithmetic, introducing new pre-solve strategies, enhanced ordering control, and a suite of auxiliary tools that streamline reduction workflows. Within {\tt FIRE7} one can completely automitize the reduction and reconstruction process with modular arithmetic on supercomuters while still keeping the original {\tt FIRE} idea that the user is supposed to provide the list of integrals requiring reduction and not care about levels of integrals being seeded.

{\tt FIRE7} extends the command-line toolset with utilities for direct manipulation of reduction tables with everything being implemented in {\tt C++}:
\begin{itemize}
  \item {\tt reconstruct} --- runs rational or modular reconstruction with different methods;
  \item {\tt tables2rules} --- converts reduction tables into Mathematica-readable rule format without invoking Mathematica;
  \item {\tt add} --- merges tables from parallel or split-master runs and supports subtraction mode for consistency checks;
  \item {\tt combine} --- reduces linear combinations of integrals using precomputed tables, useful for form-factor and amplitude applications;
  \item {\tt substitute} --- replaces analytic variables by numerical values (modular probes), ensuring full consistency with direct modular runs;
  \item {\tt diff} --- efficiently compares tables to verify identical results across different configurations or numerical probes.
\end{itemize}

Together, these utilities make it possible to conduct complete modular reductions, merging, validation, and reconstruction workflows entirely from the command line, independent of {\tt Mathematica}.
This separation of tasks facilitates large-scale, automated reductions on high-performance computing systems.

Aside of that {\tt FIRE7} comes with a bunch of new improvements:
\begin{itemize}
  \item two-stage Gaussian elimination pre-solve before seeding sampling points;
  \item configurable integral orderings (like in {\tt Kira});
  \item multiprime modular mode for parallel reconstruction allowing to run reduction in multiple prime points at the same time;
  \item split-master reduction and table combination workflow;
  \item extended support for {\tt LiteRed2} integration.
\end{itemize}  
  
\section{Conclusion}
 
Notably, various private algorithms and codes have been, and continue to be, used in the field.
Let us mention some of them because references to corresponding private versions can be found in some papers. These are, for example, {\tt IdSolver} by Czakon and {\tt Crusher} by Marquard and Seidel. 
We cannot describe them in detail in this review, as they were never made available to a broad audience; consequently, their key features remain unknown.
Let us also point out that the use of modular arithmetic for solving IBP relations was originally advocated by von Manteuffel and Schabinger \cite{vonManteuffel:2014ixa}.
The corresponding private code by von Manteuffel is called {\tt Finred}.
It is also worth mentioning methods based on syzygy equations, which can be considered a modern alternative to Gr\"obner basis techniques and the {\tt LiteRed} approach. This is because they aim to generate better relations before substituting indices.
We did not write a section on these methods since such approaches have not yet resulted in public standalone reduction programs. 
Nevertheless, a program based on this approach, {\tt Neat IBP} \cite{Wu:2023upw}, was recently integrated into the {\tt Kira} framework \cite{Wu:2025aeg} to serve as a generator of improved equations.
 
Another interesting modern approach involves improving how the sampling points are seeded for the reduction. A package that performs this task was published last year \cite{Guan:2024byi}. 
  
To summarize, we described known public general algorithms and computer codes for solving IBP relations for Feynman integrals. 
Nevertheless, even after two decades of algorithmic development for Feynman integral reduction, new ideas and methods must still be implemented to achieve optimal performance.
We therefore hope that competition in this field continues, leading to the development and practical application of new powerful algorithms and improved versions of existing ones.  
 
\section{Acknowledgment}
This work was supported by the Moscow Center for Fundamental and Applied Mathematics of Lomonosov Moscow State University under Agreement No. 075--15--2025--345.

\end{document}